# Improving Web Database Access Using Decision Diagrams[*]

Denis V. Popel[†], Nawar Al Hakeem
*Department of Computer Science,
University of Wollongong, Dubai Campus,
P.O. Box 20183, Dubai, U.A.E.
popel@ieee.org*

## Abstract

*In some areas of management and commerce, especially in Electronic commerce (E-commerce), that are accelerated by advances in Web technologies, it is essential to support the decision making process using formal methods. Among the problems of E-commerce applications: reducing the time of data access so that huge databases can be searched quickly; decreasing the cost of database design ... etc. We present the application of Decision Diagrams design using Information Theory approach to improve database access speeds. We show that such utilization provides systematic and visual ways of applying Decision Making methods to simplify complex Web engineering problems.*

## 1 Introduction

In this paper, we present the application of Decision Making methods to solve the problem of optimizing database access. At present, developments in Decision Making and Logic Design present new opportunities to provide database designers with computer-generated representations of their problems [1, 4]. Effective use of these capabilities requires managing how information is extracted from databases and using visual displays in order to enhance human performance in design tasks. Research on data representations is fundamental to the progress in optimization of interactive database applications [2].

Database access optimizers are the great tools of modern Web services to achieve high performance. Such an optimizer chooses an optimal strategy for queries processing from alternative ones. Commercial database systems have incorporated access optimizers in the last decade [6]. However, new interest in optimal sequence of queries for knowledge discovery, on-line interactive services and complex multi-media objects has caused renewed research in optimization. Such database access optimizers have been proved inadequate to the needs of these applications [5, 8].

The user interacting with an E-commerce application has a number of alternatives of which one must be chosen. The objective is to choose the best alternative (product/service) as a result of a sequence of decisions [7]. When a situation requires a series of decisions, a decision table approach cannot accommodate the multiple layers of decision making. Thus, a graph-based approach is needed. Decision Trees (DTs) and its extension Decision Diagrams (DDs) can describe these situations and add structure to the problem. DDs require less memory for representation than DTs since the DD is a reduced DT [1, 10]. DDs provide an effective method of decision making because they: layout clearly the problem so that all choices can be viewed, discussed and challenged; provide a framework to quantify the values of outcomes.

Most of the tools of modern research in optimization of Web database access - not only querying theory but also DTs, DDs and other widely used techniques - use the assumption of maximizing the achievement of some goal under specified constraints, and presume that all alternatives are known [3]. These tools have proven their usefulness in a wide variety of applications. We consider DD representation of a Web-linked database using Information Theoretic approach to minimize the uncertainty through optimization which becomes a proper heuristic to extract knowledge from the Web [2]. Our previous results explore utilization of DTs for optimizing interactive network services [9].

The rest of the paper is organized as follows. In Section 2, we review database notation, introduce basic terminology, and state the key assumptions of our work. In Section 3, we describe DDs and information theory concepts, show the relation between DDs and database information. Then, we describe the algorithms to optimize database access in Section 4. In Section 5, we present case-study results

[*]Support from the University of Wollongong (AUSTRALIA) is appreciated

[†]Support from State University of Informatics and Radioelectronics, Minsk (BELARUS), Technical University of Szczecin (POLAND), and the Fund of Fundamental Researches (BELARUS) is acknowledged

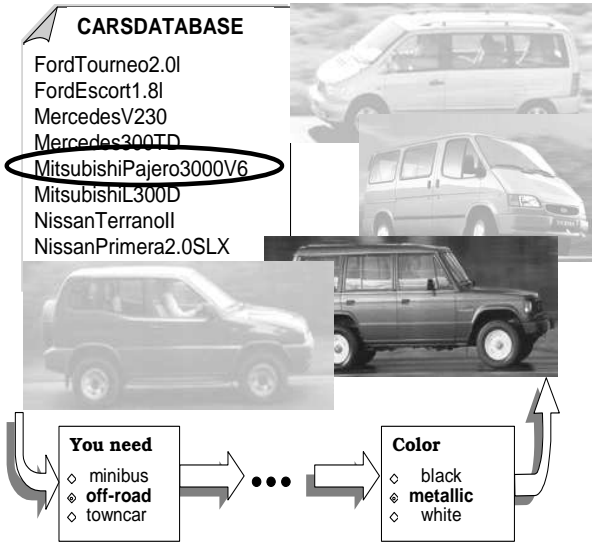

**Figure 1. An example of** cars database

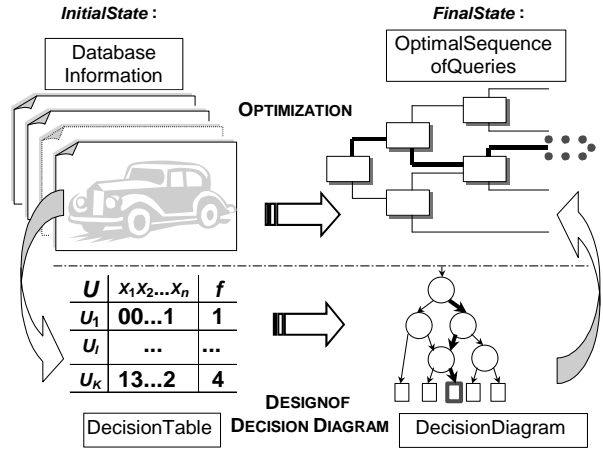

**Figure 2. Database access optimization as DT design**

on decision making benchmarks and conclude in Section 6.

## 2 Database Records and Queries

The Web E-commerce applications are based on interactive queries that explore certain products stored in a Web-linked database information. Two different principles are used when producing a query: (i) generating the queries with difficult and sometimes non-trivial questions that take a lot of time to answer; (ii) generating simple queries that contain questions with possible alternative answers. We employ the second principle and propose a new optimization strategy to achieve further performance during the execution of queries. Determining the optimal sequence can be solved using DDs and information theoretic measures.

**Example 1.** *To illustrate optimization of Web database access, the following example of Internet Shopping is used. We have chosen the **demo cars database**[1] to be used by a hypothetical company that sells cars on the Web (Figure 1).*

Our approach is based on converting a Web-linked database to canonical form such as decision table and representation of queries as DD structure (Figure 2). Table 1 shows the correspondence of the terminology for the DD elements.

The underlying approach typically involves variables (features), $x$, and response, $f$. In the following, we consider the $m$-valued logic function $f: \mathbf{A}^n \to \mathbf{B}$ over the variable set $X = \{x_1, \cdots, x_n\}$, where $\mathbf{A}=\{0, r-1\}$ and $\mathbf{B}=\{0, m-1\}$. Here, $n$ is the number of $r$-valued variables.

[1] http://www.elshopsoft.com/download/samples/

**Example 2.** *The company from Example 1 sells the following car modifications:*

$f=0$: *Ford Tourneo 2.0l - minibus, petrol engine, manual gear, velour interior, controllable catalizator, fuel spent 10.9, price 28,900;*

$f=1$: *Ford Escort 1.8l - town car, diesel or diesel/turbo engine, manual gear, cherry color, velour or leather interior, fuel spent 6.4, price 19,900 - 20,300;*

$f=2$: *Mercedes V 230 - minibus, petrol or petrol/turbo engine, automatic gear, white color, velour interior, controllable catalizator, fuel spent 11.6, price 36,600;*

$f=3$: *Mercedes 300TD - town car, diesel engine, automatic gear, white color, velour interior, fuel spent 8.4, price 27,500;*

$f=4$: *Mitsubishi Pajero 3000 V6 - off-road, desiel/turbo engine, manual or automatic gear, white color, leather interior, fuel spent 13.7, price 24,800 - 25,600;*

$f=5$: *Mitsubishi L300D - minibus, diesel engine, manual gear, metallic color, leather interior, fuel spent 9.8, price 25,700;*

$f=6$: *Nissan Terrano II - off-road, petrol engine, manual gear, metallic color, velour or leather interior, controllable catalizator, fuel spent 11.1, price 24,600;*

$f=7$: *Nissan Primera 2.0SLX - town car, diesel or diesel/turbo engine, automatic gear, velour interior, fuel spent 7.9 - 8.2, price 18,350;*

*Characteristics of cars described by the multiple-valued variables:*

$x_1$: *catalizator - none (0), controllable (1);*

$x_2$: *color - black (0), cherry (1), metallic (2), white (3);*

$x_3$: *engine - petrol (0), diesel (1), petrol/turbo(2), diesel/turbo (3);*

$x_4$: *interior - leather (0), velour (1);*

$x_5$: *gear - manual (0), automatic (1);*

## Table 1. Terminology relationship between logic and database functions

| Logic Function | Database |
|---|---|
| Variable $x$ | Characteristic of the product |
| Function $f$ | Range of the proposed products |
| Variable value $x = a$ | An alternative |
| Function value $f = b$ | Product identifier |

$x_6$: fuel spent - less than 8.0 (0), between 8.0 and 10.0 (1), between 10.0 and 12.0 (2), greater than 12.0 (3);

$x_7$: price - less than 20,000 (0), between 20,000 and 25,000 (1), between 25,000 and 30,000 (2), greater than 30,000 (3);

$x_8$: purpose - minibus (0), off-road (1), town car (2).

## 3 Database and Logic Function

### 3.1 Database Decomposition

Let us investigate the decomposition of database information. This can be represented as decomposition of logic function $f$ with respect to variable $x$ into uniquely determined sub-functions so that it is possible to reconstruct $f$ if the sub-functions are known. For a logic function $f$, $f_c = f_{|x_i=c} = f(x_1, \ldots, x_{i-1}, c, x_{i+1}, \ldots, x_n)$ is called a *cofactor* or sub-function of $f$, when $x$ is fixed to $c \in \{0, \ldots, r\}$.

**Definition 1.** *A Decomposition of a function $f$ is defined as $f = Decomposition(x, f_0, \ldots, f_r)$, such that for $\forall x \in X$, there exist $r$ uniquely determined cofactors $f_0, \ldots, f_{r-1}$.*

### 3.2 Representation of Logic Functions

Any logic function $f$ can be uniquely determined by a *truth table* on $k$ combinations of variable values. In decision making applications, the term *decision table* is used instead of truth table.

**Example 3.** *The decision table for the database from Example 1 is given in Table 2 ($k = 19$).*

#### 3.2.1 Decision Diagrams and Graph-Based Notations

Decision Trees (DTs) and Decision Diagrams (DDs) are graph-based structures which have become the advanced structures in Logic Design and Decision Making for representing and manipulating functions and discrete data

## Table 2. Truth table of logic function $f$ from Example 1

| Model | $x_1$ | $x_2$ | $x_3$ | $x_4$ | $x_5$ | $x_6$ | $x_7$ | $x_8$ | $f$ |
|---|---|---|---|---|---|---|---|---|---|
| Ford Tourneo 2,0l | 1 | 0 | 0 | 1 | 0 | 2 | 2 | 0 | 0 |
| | 1 | 1 | 0 | 1 | 0 | 2 | 2 | 0 | 0 |
| | 1 | 2 | 0 | 1 | 0 | 2 | 2 | 0 | 0 |
| | 1 | 3 | 0 | 1 | 0 | 2 | 2 | 0 | 0 |
| Ford Escort 1,8 | 0 | 1 | 1 | 1 | 0 | 0 | 0 | 2 | 1 |
| | 0 | 1 | 3 | 1 | 1 | 1 | 1 | 2 | 1 |
| Mercedes V230 | 1 | 3 | 0 | 0 | 1 | 2 | 3 | 0 | 2 |
| | 1 | 3 | 0 | 1 | 1 | 2 | 3 | 0 | 2 |
| | 1 | 3 | 2 | 1 | 1 | 2 | 3 | 0 | 2 |
| Mercedes 300TD | 0 | 3 | 1 | 1 | 1 | 1 | 2 | 2 | 3 |
| Mitsubushi Pajero 3000 V6 | 0 | 3 | 3 | 0 | 0 | 3 | 1 | 1 | 4 |
| | 0 | 3 | 3 | 0 | 1 | 3 | 2 | 1 | 4 |
| Mitsubishi L300D | 0 | 2 | 1 | 0 | 0 | 1 | 2 | 0 | 5 |
| Nissan Terrano II | 1 | 2 | 0 | 0 | 0 | 2 | 1 | 1 | 6 |
| | 1 | 2 | 0 | 1 | 0 | 2 | 1 | 1 | 6 |
| Nissan Primera 2,0SLX | 0 | 0 | 1 | 1 | 1 | 0 | 0 | 2 | 7 |
| | 0 | 1 | 3 | 1 | 1 | 1 | 0 | 2 | 7 |
| | 0 | 2 | 3 | 1 | 1 | 1 | 0 | 2 | 7 |
| | 0 | 3 | 3 | 1 | 1 | 1 | 0 | 2 | 7 |

[1, 10]. The core of the data structures is a directed acyclic graph which forms a canonical representation of a given function.

**Definition 2.** *Decision Diagram is a connected directed acyclic graph with vertex (node) set and edge set where:*

**(i)** *Each non-terminal vertex is labeled by a variable $x$ and assigned as a decision variable. Also, such a vertex corresponds to a decomposition step of the function $f$ into sub-functions (outgoing edges: $edge, \ldots, edge_r$) with respect to the variable $x$.*

**(ii)** *A terminal vertex is labeled with the leaf value and has no successors. But a non-terminal vertex has exactly $r$ successors for $r$-valued variable.*

**(iii)** *When reduction is performed: (i) any node with identical successors DD1, ..., DDr is removed (Figure 3(a)); (ii) two nodes with isomorphic DDs are merged (Figure 3(b)).*

**(iv)** *A DD is called ordered if the variable $x$ appears in the same order in each path from the root to a terminal vertex. A DD is called free if the order of variables $x$ is free along with each path from the root to a terminal vertex. In other words, the term 'free' means that different variables and expansion types can occur at every level of DD.*

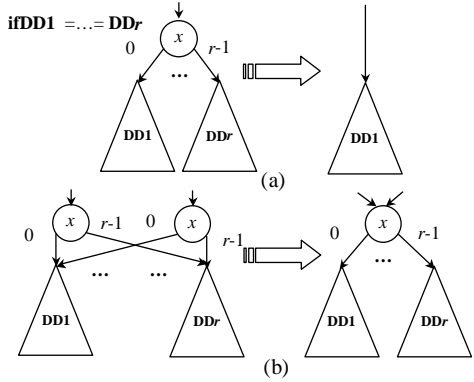

**Figure 3. Reduction rules for DD design**

Free DDs allow more efficient representation while keeping (nearly) all the properties of ordered DDs [4]. We deal with free DDs only so the term 'free' will be omitted.

**Example 4.** *The graph in Figure 4(a) represents an ordered DD for the function f (Example 1). The DD in Figure 4(b) is free. The effect of reducing the number of DD nodes is demonstrated.*

#### 3.2.2 Relation Between Decision Diagrams and Database Information

A DT (or generally DD) is a chronological representation of the decision process by a network that utilizes two types of nodes: decision nodes, represented by *choice* nodes (values of a function), and *chance* nodes (variables). Constructing a DD requires building a logical structure for the problem. Here is a sketch description of how to design a DD: 1. draw the DT using choice nodes to represent decisions, and chance nodes to represent uncertainty states; 2. evaluate the DT to make sure all possible outcomes are included; 3. reduce DT to DD.

We can determine the best decision from the graph by starting from the root and going forward. From the above graph our decision is as follows: 1. ask the consumer several questions to discover his interest; 2a. if answers lead to a particular product, then select the product (final decision); 2b. otherwise repeat questions.

**Definition 3.** *$i$-path is a path from the root of DD to a terminal node assigned with logic value $i$. x-path is a path from the root of DD to a terminal node assigned with no value.*

Each $i$-path defines a sub-set of variable's values that uniquely correspond to a record in the initial database [9].

**Example 5.** *Let us consider the function $f$ given by DD (Figure 4(b)). Its path in bold corresponds to target $f = 7$ (7-path). It means that during the Internet Shopping we will follow this path and choose Nissan Primera 2.0SLX.*

The major problem is to choose the variable for DD design that will optimize Web database access by minimizing levels of DD for quick search and reducing size of DD for efficient memory allocation. This problem can be solved using information theoretic measures as optimization criteria.

### 3.3 Information Theory and Optimization

In order to quantify the content of information for a finite field of events $A = \{a_1, a_2, \cdots, a_n\}$ with probabilities distribution $\{p(a_i)\}$, $i = 1, 2, \cdots, n$, Shannon introduced the concept of entropy [11]. *Entropy* of the finite field $A$ is given by (logarithm is base 2)

$$H(A) = -\sum_{i=1}^{n} p(a_i) \cdot \log p(a_i), \qquad (1)$$

Suppose, there are two finite fields of events $A$ and $B$ with probability distribution $\{p(a_i)\}, i = 1, 2, \cdots, n$, and $\{p(b_j)\}, j = 1, 2, \cdots, m$, respectively. Let $p(a_i, b_j)$ be the probability of the joint occurrence of $a_i$ and $b_j$. For any particular value $a_i$, that $A$ can assume, there is a conditional probability $p(a_i|b_j)$ that $B$ has a value $b_j$. It is expressed by $p(a_i|b_j) = \frac{p(a_i, b_j)}{\sum_{i=1}^{n} p(a_i, b_j)}$. The *conditional entropy* of $A$ given $B$ is defined by

$$H(A|B) = -\sum_{i=1}^{n}\sum_{j=1}^{m} p(a_i, b_j) \cdot \log p(a_i|b_j). \qquad (2)$$

Here, we deal with two finite fields: set of values of function $f$ and set of values of variable $x$. We calculate the probability $p_{|f=b} = k_{|f=b}/k$, where $k_{|f=b}$ is the number of assignments of values to variables (patterns) for which $f = b$ and $k$ is the total number of assignments. Other probabilities are calculated in the same way.

**Example 6.** *Consider the function $f$ from Example 1. The probabilities of its output values are $p_{|f=0} = p_{|f=7} = {}^4/_{19}$, $p_{|f=1} = p_{|f=4} = p_{|f=6} = {}^2/_{19}$ and $p_{|f=2} = p_{|f=3} = p_{|f=5} = {}^1/_{19}$. The entropy of the function is $H(f) = -2 \cdot {}^4/_{19} \cdot \log_2 {}^4/_{19} - 3 \cdot {}^2/_{19} \cdot \log_2 {}^2/_{19} - 3 \cdot {}^1/_{19} \cdot \log_2 {}^1/_{19} = 2.64$ bit. The conditional entropy of the function $f$ with respect to variable $x_1$ is $H(f|x_1) = {}^9/_{19} \cdot 1.24 + {}^{10}/_{19} \cdot 1.61 = 1.43$ bit, and also $H(f|x_2) = 1.08$ bit, $H(f|x_3) = 1.00$ bit, $H(f|x_4) = 1.80$ bit, $H(f|x_5) = 1.53$ bit, $H(f|x_6) = 1.01$ bit, $H(f|x_7) = 0.84$ bit, $H(f|x_8) = 0.99$ bit.*

We utilize the presented information theoretic measures for optimization of database access. The *criterion* to choose

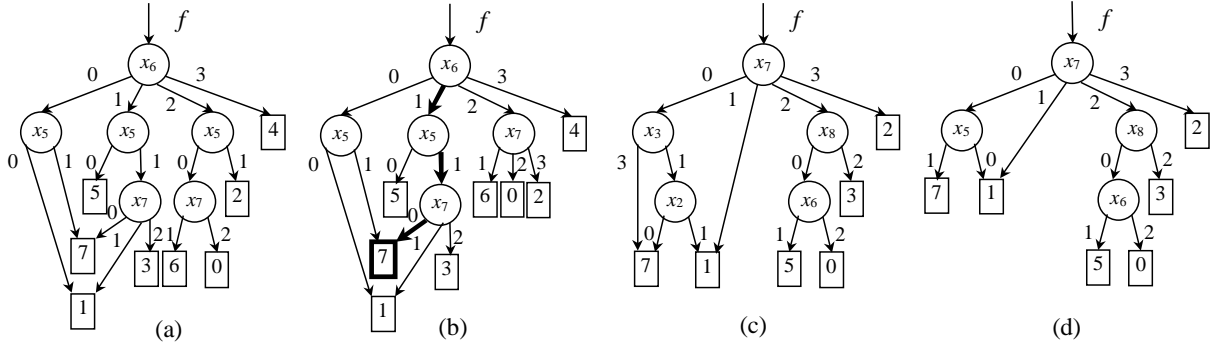

**Figure 4. Samples of (a) ordered and (b) free DDs, and resultant DDs produced (c) by greedy algorithm $Info_{Greedy}$ and (d) by iteration algorithm $Info_{Iter}$ (second iteration) for the function $f$ from Example 1**

a decomposition variable $x$ for the arbitrary level of DD is that the conditional entropy of the function with respect to this variable has to be minimal:

$$H(f|x) = min(H(f|x_i) \mid \forall \ x_i). \quad (3)$$

As a measure of cost, we use the number of levels and the number of nodes in the final DD. This choice is motivated by the major optimization objectives in Internet Shopping, related to reduction of number of queries and overall memory size for DD allocation.

The main reasons for using information theoretic measures to optimize data access are:

1. The behaviour of entropy function is close to the behaviour of such parameters as the number of nodes and the number of levels in DD. The results from [9] show the dependence of the number of nodes in DT expression upon the entropy function.

2. The choice in each particular case is mainly justified by the *uncertainty* of decision making whose estimation is closely related to entropy measures. This implies that the sequence with less uncertainty (DDs) should be designed taking into consideration the entropy criterion.

3. The results of optimization are very sensitive to variable ordering, e.g. the number of nodes may vary from linear to exponential [10].

Next, we present a simple example to compare a classical method and an entropy-based method of variable ordering.

## 4 Algorithms to Optimize Database Access

Generally, our algorithm to optimize database access performs as follows,

⋄ Initially, a canonical representation, i.e. truth table, is generated for the given database information as described in Section 3.

⋄ $Info_{Greedy}$ (greedy strategy) or $Info_{Iter}$ (iteration strategy) algorithm is applied. The nodes of the DD are assigned by variables in accordance with the information theoretic criterion (Equation (3)). The DT is optimized via reduction of the number of nodes.

⋄ The sequence of queries is formed according to the constructed DD.

### 4.1 Greedy Strategy - Simple Case

First, we describe a greedy algorithm to optimize database access according to an information criterion. A sketch of the algorithm is given in Figure 5.

The basic idea here is that we employ recursion when constructing DDs. The ordering restriction is relaxed, i.e., (i) each variable appears once on each path and (ii) the order of variables along each path may be different [4]. Our greedy algorithm for logic functions minimization is:

**Stage 1.** At each step of DD design, i.e. attaching a current node to the DD, the information theoretic measures for decomposition are calculated for each variable.

**Stage 2.** The variable $x$, that corresponds to minimal $H(f|x)$, is assigned to the current DD node.

**Stage 3.** Sub-DDs for the sub-functions (outgoing edges of current DD node) obtained by decomposition with respect to variable $x$ are recursively constructed.

**Stage 4.** Algorithm terminates if the leaves are archived for each sub-DD (DD is completed) for the given logic function $f$.

```
Input Logic function f
Output DD - Decision Diagram

Info_Greedy(f)
{ if(f = c, where c = const) then {
       DD ← leaf(c); return;
  }
  for(∀x_i)
      Calculate information measures H(f|x_i)
  Choose variable x where:
              H(f|x) = min(H(f|x_i) | ∀x_i);
  Attach node assigned by variable x to DD
              DD ← node(x);
  for(∀f_s of decomposition given variable x)
      Recursively construct the sub-DDs DD_s:
              DD_s = Info_Greedy(f_s);
  return;
}
```

**Figure 5. Sketch of the $Info$ algorithm to realize greedy strategy**

```
Input Logic function f, number of iterations Iter
Output DD - Decision Diagram

for(iter = 1; iter ≤ Iter; iter + +) {
Info_Iter(f) {if(f = c, where c = const) then {
       DD ← leaf(c); return;
  }
  for(∀x_i)
      Calculate information measures H(f|x_i)
  Range the variables x_i by increasing H(f|x_i)
  Choose variable x from list of ranging couples
  Attach node assigned by variable x to DD
              DD ← node(x);
  for(∀f_s of decomposition given variable x)
      Recursively construct the sub-DDs DD_s:
              DD_s = Info_Iter(f_s);
  return;
}
  Store minimal DD, according to cost criterion
}
```

**Figure 6. Sketch of the $Info$ algorithm to realize iteration strategy**

The obtained DD is shown in Figure 4(c). The number of non-terminal nodes is four and the maximum number of levels is three.

### 4.2 Iteration Strategy

We present below an extension of the greedy algorithm that can be used in practical applications. A concept of ranging variables $x_1, \ldots, x_n$ using information theoretic criterion is supposed to improve the characteristics of greedy strategy and optimize Web access.

1. During information measures calculation, we store the list of variables $x$ ranging by increasing $H(f|x)$: $x_{j_1}, ... x_{j_t}, ..., x_{j_n}$, so that $H(f|x_{j_t}) < H(f|x_{j_{t+1}})$ (Figure 7(b)), in contradiction to lexicographical (naive) order (Figure 7(a)).

2. At each iteration, we choose the variable from the list $x_{j_1}, ..., x_{j_t}, ..., x_{j_n}$, corresponding to the current iteration.

We add the number of iterations as input data for the extended algorithm (Figure 6). Such an improvement of the basic algorithm does not guarantee the minimal solution, but near the minimal one. It is easy to show that algorithm $Info_{Iter}$, with parameter $Iter = 1$, realizes the greedy strategy. We can obtain the results that will be near the exact ones by increasing the number of iterations:

Algorithm   $Info_{Greedy}$   ⟵ $Info_{Iter}$ ⟶   Exact
$Iter =$          1         ... 10 ... 100 ...      ...

**Example 7.** *Let us consider how the algorithm $Info_{Iter}$ runs for the function $f$ from Example 1. At the second iteration, we obtain DD (Figure 4(d)). We can conclude that*

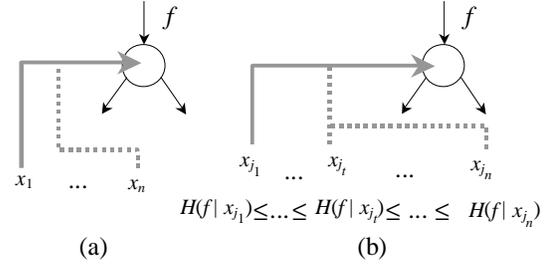

**Figure 7. Lexicographical (naive) order and the order based on ranging the variables in accordance with their information measures**

*three requests will be enough to explore all car modifications proposed by the company. Firstly, we should generate a query that contains a question about customers' pay abilities ($x_7$), then either gear preferences ($x_5$) or car purpose ($x_8$), and fuel spent ($x_6$).*

## 5 Experiments and Practical Benefits

In the first series of experiments with algorithms $Info_{Greedy}$ and $Info_{Iter}$ for decision making, Machine Learning benchmarks were used (Table 3). In this table, $N/level/t$ means the number of DT/DD nodes, the number of DT/DD levels and run-time in CPU seconds (Pentium III 650Mhz, 48Mb). We state $Iter = 10$ for $Info_{Iter}$.

**Observation 1.** *$Info_{Iter}$ algorithm produces DTs with about 10% fewer nodes (DDs with about 7% fewer nodes) and about 12% fewer levels (9% fewer levels for DDs) than*

**Table 3. Results of $Info_{Greedy}$ and $Info_{Iter}$ in decision making applications**

|  |  |  | $Info_{Greedy}$ | | $Info_{Iter}$ | |
|---|---|---|---|---|---|---|
|  |  |  | DT | DD | DT | DD |
|  | $r = m$ | $k$ | N/level/t | N/level/t | N/level/t | N/level/t |
| shuttle | 4 | 1695 | 740/6/8.31 | 651/6/10.25 | 740/6/8.31 | 651/6/10.25 |
| monks1te | 4 | 432 | 10/3/0.26 | 10/3/0.26 | 10/3/0.26 | 10/3/0.26 |
| monks1tr | 4 | 124 | 17/5/0.05 | 15/5/0.19 | 13/3/0.24 | 11/3/1.84 |
| monks2te | 4 | 432 | 10/3/0.26 | 10/3/0.26 | 10/3/0.26 | 10/3/0.26 |
| monks2tr | 4 | 169 | 85/6/0.02 | 78/6/0.12 | 79/6/0.55 | 71/6/1.13 |
| monks3te | 4 | 432 | 73/5/0.56 | 36/4/2.88 | 5/3/1.68 | 5/3/1.68 |
| monks3tr | 4 | 122 | 39/5/0.07 | 32/5/0.75 | 22/5/0.62 | 19/5/2.39 |
| **Total** |  |  | 974/33/9.53 | **832/32**/14.71 | *879/29*/11.92 | **777/29**/17.81 |

$Info_{Greedy}$ does.

**Observation 2.** $Info_{Iter}$ with DDs output gives about 12% fewer nodes than $Info_{Iter}$ with DTs output.

In the second series of experiments, we tested the proposed algorithms for different Internet market examples. The optimization results provide more friendly and faster user interactions.

Possible benefits from using DDs are: more compact database representations and faster access, better optimization using different criteria (DD size, levels' no.), and flexibility in developing and updating electronic catalogs. The application of Internet shopping is an example where Web site customers will be able to buy a product using intuitive navigation due to DDs since hierarchical data representation is similar to the way decisions are made.

## 6 Concluding Remarks

In this paper, we addressed the problem of optimizing database access by using hierarchical organization of database information. We have developed computer-aided support for the easy construction of DDs and DTs. The optimization methods using graph-based structures have found wide applicability not only in E-commerce applications but also in logic design, computer-aided diagnostics in medicine, and other decision making problems [12].

The algorithms produce efficient DDs using information theoretic approach. They provide significant improvement in the number of queries needed to extract information from a database. The experimental results are encouraging and the algorithms are easy to construct.